\documentclass[12pt,preprint]{emulateapj}

\shorttitle{GRB 120422A/SN 2012bz Host Properties}
\shortauthors{Levesque et al.}

\begin{document}

\title{Host Galaxy Properties of the Subluminous GRB 120422A/SN 2012bz\footnotemark[1]} \footnotetext[1]{This paper includes data gathered with the 6.5 meter Magellan Telescopes
located at Las Campanas Observatory, Chile.}
\author{Emily M. Levesque$^{2,3}$, Ryan Chornock$^4$, Alicia M. Soderberg$^4$, Edo Berger$^4$, and Ragnhild Lunnan$^4$}

\begin{abstract}
GRB 120422A is a nearby ($z = 0.283$) long-duration GRB (LGRB) detected by Swift with $E_{\gamma,iso} \sim 4.5 \times 10^{49}$ erg. It is also associated with the spectroscopically-confirmed broad-lined Type Ic SN 2012bz. These properties establish GRB 120422A/SN 2012bz as the sixth and newest member of the class of subluminous GRB/SNe. Observations also show that GRB 120422A/SN 2012bz occurred at an unusually large offset ($\sim$8 kpc) from the host galaxy nucleus, setting it apart from other nearby LGRBs and leading to speculation that the host environment may have undergone prior interaction activity. Here we present spectroscopic observations using the 6.5m Magellan telescope at Las Campanas. We extract spectra at three specific locations within the GRB/SN host galaxy, including the host nucleus, the explosion site, and the ``bridge" of diffuse emission connecting these two regions. We measure a metallicity of log(O/H) + 12 = 8.3 $\pm$ 0.1 and a star formation rate per unit area of 0.08 M$_{\odot}$ yr$^{-1}$ kpc$^{-2}$ at the host nucleus. At the GRB/SN explosion site we measure a comparable metallicity of log(O/H) + 12 = 8.2 $\pm$ 0.1, but find a much lower star formation rate per unit area of 0.01 M$_{\odot}$ yr$^{-1}$ kpc$^{-2}$. We also compare the host galaxy of this event to the hosts of other LGRBs, including samples of subluminous LGRBs and cosmological LGRBs, and find no systematic metallicity difference between the environments of these different subtypes.
\end{abstract}

\section{Introduction}
\footnotetext[2]{CASA, Department of Astrophysical and Planetary Sciences, University of Colorado 389-UCB, Boulder, CO 80309, USA; \texttt{Emily.Levesque@colorado.edu}}
\footnotetext[3]{Einstein Fellow}
\footnotetext[4]{Harvard-Smithsonian Center for Astrophysics, 60 Garden St., Cambridge, MA 02138, USA}

Recent work on LGRBs at $z < 1$ has suggested a connection between LGRBs and low-metallicity host environments. Their host galaxies, on average, fall below the luminosity-metallicity and mass-metallicity relations for star-forming galaxies out to $z \sim 1$ (e.g. Stanek et al.\ 2006; Levesque et al.\ 2010a,b; Mannucci et al.\ 2011). However, the physical mechanism driving this apparent metallicity trend is still poorly understood. LGRBs do not appear to be exclusive to low-metallicity environments, with several super-solar host galaxies and explosion sites for LGRBs (e.g. Levesque et al.\ 2010b,c). There is also no apparent correlation between host metallicity and gamma-ray energy release for LGRBs (Levesque et al.\ 2010e), a result that is at odds with previous predictions of LGRB progenitor models (MacFadyen \& Woosley 1999).

However, it is possible that our current picture of these objects is oversimplified. There appears to be evidence for multiple sub-classes of LGRBs. Detailed studies of nearby LGRBs have revealed a subset of these events with unusually low gamma-ray energies and luminosities ($E_{\gamma,iso} \lesssim 10^{50}$ erg and $L \lesssim 10^{49}$ ergs s$^{-1}$, e.g. Kulkarni et al.\ 1998, Soderberg et al.\ 2006a, Guetta \& Della Valle 2007). These subluminous events, which dominate the $z \lesssim 0.3$ LGRB population, are thought to be much more frequent that the higher-luminosity ($E_{\gamma,iso}$ $\sim 10^{52}$) cosmological LGRBs detected at higher redshifts. Each subluminous LGRB is also associated with a spectroscopically identified supernova (see Woosley \& Bloom 2006 for a review). However, supernova associations are not restricted to only subluminous LGRBs: GRB 030329/SN 2003dh ($z = 0.168$) and GRB 091127/SN 2009nz ($z = 0.49$) are both associated with spectroscopically-confirmed Ic-BLs despite having ``cosmological" luminosities (Stanek et al.\ 2003, Berger et al.\ 2011), and a number of other more distant bursts have shown late-time photometric rebrightenings in their afterglow lightcurves from associated SNe (e.g. Bloom et al.\ 2002a, Soderberg et al.\ 2005, 2006b; Cano et al.\ 2011). Conversely, two subluminous LGRBs (060505 and 060614)have also been observed that show {\it no} evidence of any associated supernovae (Fynbo et al.\ 2006) although the classification of these bursts and their connection to the general LGRB population is still uncertain (e.g. Gal-Yam et al.\ 2006, Ofek et al.\ 2007, Zhang et al.\ 2007, Th\"{o}ne et al.\ 2008). It is currently unclear whether subluminous LGRBs represent a phenomenologically-distinct subclass within the broader LGRB sample (see Cobb et al.\ 2006, Zhang et al.\ 2012).

The sixth and newest member of this potential subclass of subluminous GRB/SNe, GRB 120422A, was detected by the {\it Swift} Burst Alert Telescope on 12 April 22 at 07:12:03 UT (Troja et al.\ 2012). Prompt emission observations determined a duration of $T_{90} \sim 5$ s, while early follow-up observations measured a redshift of $z = 0.283$ based on Mg II absorption in the optical afterglow of the GRB as well as nebular emission features from the presumed host galaxy, SDSS J090738.51+140108.3 (Schulze et al.\ 2012, Tanvir et al.\ 2012).  Subsequently an associated Ic-BL supernova, SN 2012bz, was spectroscopically confirmed by Wiersema et al.\ (2012) and found to be very similar to other Ic-BLs associated with LGRBs (Melandri et al.\ 2012). The total isotropic energy of the burst was measured to be $E_{\gamma, iso} \sim 4.5 \times 10^{49}$ erg, with a peak energy of $\sim$53 keV, marking it as subluminous compared to the general LGRB population (Schulze et al.\ 2012, Zhang et al.\ 2012).

GRB 120422A/SN 2012bz is unique among nearby LGRBs due to its localization at an unusually large offset from the center of its host galaxy - Tanvir et al.\ (2002) measure a projected physical offset of $\sim$8 kpc, much larger than the median offset measured in the sample of Bloom et al.\ (2002b). Such an offset is one of the largest observed for an LGRB, which are typically localized in the brightest and bluest regions of their hosts (Bloom et al.\ 2002b, Fruchter et al.\ 2006). This suggests that GRB 120422A occurred in a star-forming region near the outskirts of the host, similar to other events such as GRB 980425 and GRB 990705 (e.g. Christensen et al.\ 2008, Bloom et al.\ 2002b); however, the absence of clearly identified spiral arms in this host has led to speculation that the star-forming region hosting this burst may have been produced by an interacting system (Tanvir er al.\ 2012, Perley et al.\ 2012, Sanchez-Ramirez et al.\ 2012). 

Here we present spectroscopy of several locations within the GRB 120422A/SN 2012bz host galaxy. We discuss the observations and describe the reduction and analysis applied to these spectra in \S2. Based on these we derive ISM properties within the host (\S3) and place GRB 120422A/SN 2012bz in context with the larger LGRB host environment population, considering the implications for our understanding of LGRBs and the subluminous LGRB subclass (\S4). Throughout this work we adopt the standard cosmological parameters $H_0=71$ km s$^{-1}$ Mpc$^{-1}$, $\Omega_m=0.27$, and $\Omega_\Lambda=0.73$.

\section{Observations and Analysis}
On 2012 May 16-17 UT we obtained two 1200-s spectra of the GRB 120422A/SN 2012bz host galaxy with the Low Dispersion Survey Spectrograph (LDSS3) on the Magellan/Clay 6.5-m telescope at Las Campanas Observatory. The 1$\arcsec$ slit was aligned along both the bright nucleus of the host galaxy and the explosion site of the GRB/SN, at a position angle of 50$^{\circ}$, an airmass of $\sim$1.5-1.6, and a seeing of $\sim$0.7$\arcsec$ It should be noted that this is $\sim$90$^{\circ}$ off from the parallactic angle of 141$^{\circ}$, leading to moderate slit losses in the bluest parts of the spectra. As a result, we also obtained a third 2100-s spectrum, centered on the GRB 120422A/SN 2012bz explosion site and oriented at the parallactic angle. Each of the spectra cover a wavelength range of $\sim$4700\AA-8700\AA, with a dispersion of 2.0\AA/pixel. During the same night we acquired a 50-s observation of the spectrophotometric standard LTT 2415 (Hamuy et al.\ 1994) as well as observations of quartz and arc lamps. As these observations were taken only $\sim$25 days after initial detection of the GRB/SN, we were able to easily localize the explosion site through identification of the SN emission. The GRB/SN Site is substantially offset from the Nucleus by $\sim$1.9$\arcsec$ ($\sim$8 kpc at $z = 0.283$); the size of the host as a whole is approximately $3\arcsec \times 1.6\arcsec$ (13 kpc $\times$ 7 kpc).

The data were reduced using standard routines in IRAF\footnotemark\footnotetext{IRAF is distributed by NOAO, which is operated by AURA, Inc., under cooperative agreement with the NSF.}, including bias correction and cosmic ray removal. We applied a flatfield correction based on internal quartz lamp flats, and subtract background skylines from the two-dimensional data to minimize residuals in the extracted spectra. Three spectra were extracted from our observations along the host galaxy using an optimal extraction algorithm; each extracted region had a width of 6 pixels ($\sim1.14\arcsec$), with deviant pixels identified and rejected based upon the assumption of a smoothly varying profile. The first spectrum was centered on the bright nucleus of the host (hereafter ``Nucleus"). The second spectrum was centered on the bridge of extended emission  (hereafter ``Bridge") to the southwest of the host nucleus (Figure 1; see also Perley et al.\ 2012), with the brighter continuum of the nucleus used as a robust trace when extracting this nearby dimmer spectrum. The third and final spectrum was centered on the explosion site of GRB 120422A/SN 2012bz (hereafter ``GRB/SN Site"), tracing the bright emission of the supernova. In addition, we used our observation at the parallactic angle to extract a second spectrum centered at the GRB/SN, and combined the two GRB/SN Site spectra to maximize the total exposure time in this key region of the host. Wavelength calibration and flux calibration were performed using arc lamp spectra and quasi-simultaneous spectrophotometric standard observations. We measured the emission line fluxes in these spectra using the IRAF task \texttt{splot} in the {\texttt{kpnoslit} package to fit Gaussians to the line profiles, with multiple Gaussians used in the case of emission lines that exhibit clear asymmetries. The fluxes measured for each emission feature are given in Table 1.

\section{Host ISM Properties}
Key diagnostic emission features detected in the spectra -- [OII] $\lambda$3727, H$\beta$, [OIII] $\lambda\lambda$4959,5007, H$\alpha$, and [NII] $\lambda$6584 -- are shown in Figure 2. While we include our spectra of the [OII] $\lambda$3727 features for comparison purposes, it should be noted that the [OII] fluxes in the Nucleus and Bridge spectra can only be considered as lower limits (a result of significant slit losses in the blue due to observing at $\sim$90$^{\circ}$ off of the parallactic angle) and are not useful for diagnostic purposes as a result. However, the differential slit losses in these spectra are small over short wavelength intervals, allowing us to accurately use diagnostics that depend on ratios of emission lines at close wavelengths (i.e., [OIII] $\lambda$5007 and H$\beta$). Across the host, the HII region emission features are notably weaker in the Bridge and GRB/SN Site spectra as compared to the Nucleus spectra. Spectra at the GRB/SN Site also show continuum contribution from the Ic-BL SN emission. The Ic-BL SN features are sufficiently broad and smooth over short wavelength intervals (see Figure 2) that we were able to adequately subtract the background by linearly interpolating across the wavelength of the emission lines to measure the fluxes.

We determined E($B-V$) in the direction of the GRB 120422A/SN 2012bz host galaxy based on the fluxes of the H$\alpha$ and H$\beta$ emission features, the Cardelli et al.\ (1981) reddening law with $R_V = 3.1$ mag, and a Balmer decrement of H$\alpha$/H$\beta = 2.87$ assuming case B recombination with an electron temperature $T_e = 10^4 K$ (Osterbrock 1989), in good agreement with average electron temperatures observed in GRBs (e.g. Hammer et al.\ 2006). Accounting for a foreground Galactic reddening of E($B-V)$ = 0.03 mag (Schlegel et al.\ 1998), we find a total line-of-sight E($B-V$) = 0.24$\pm$0.03 mag in the direction of the Nucleus, and a slightly higher E($B-V$) = 0.29$\pm$0.10 mag in the direction of the Bridge and E($B-V$) = 0.31$\pm$0.13 mag in the direction of the GRB/SN site. These values show good agreement with the broad range E($B-V$)s determined for previously-studied LGRBs hosts, which have an average of 0.37 $\pm$ 0.36 (Levesque et al.\ 2010a,b).

Metallicity values were determined based on the Pettini \& Pagel (2004) calibration of the {\it O3N2} ([NII] $\lambda$6584/H$\alpha$)/([OIII] $\lambda$5007/H$\beta$) diagnostic. For the Nucleus we determine a metallicity of log(O/H) + 12 = 8.3 $\pm$ 0.1(errors are derived from the systematic uncertainty of the diagnostic calibrations; see Kewley \& Ellison 2008). For the Bridge and GRB/SN Site we are only able to determine upper limits for the [NII] $\lambda$6584 emission feature and a resulting metallicity upper limit of log(O/H) + 12 $<$ 8.3 at both locations using the {\it O3N2} diagnostic. However, by observing at the parallactic angle for the GRB/SN Site, we are also able to use the [OII] $\lambda$3727 flux to estimate metallicity using the $R_{23}$ diagnostic calibration from Kobulnicky \& Kewley (2004). We find that the GRB/SN Site lies on the ``turn-over" of this double-valued diagnostic, corresponding to a metallicity of log(O/H) + 12 = 8.4 $\pm$ 0.1. Applying the conversion between the $R_{23}$ and {\it O3N2} diagnostics from Kewley \& Ellison (2008), this corresponds to an {\it O3N2} metallicity of log(O/H) + 12 = 8.2 $\pm$ 0.1, in agreement with the Nucleus metallicity to within the systematic errors of the two metallicity diagnostics.

We use the flux of the H$\alpha$ features in our data to determine SFR measurements at the Nucleus, Bridge, and GRB/SN Site based on the relation of Kennicutt (1998). With a 1$\arcsec$ slit and extraction apertures of 1.14$\arcsec$ (corresponding to six pixels and a pixel scale of 0.19$\arcsec$ for LDSS3), these SFRs correspond to an area of $\sim$ 4.3 kpc $\times$ 4.9 kpc, or $\sim$21 kpc$^2$ at $z = 0.293$. We measure SFRs of $\sim$0.08 M$_{\odot}$ yr$^{-1}$ kpc$^{-2}$ in the Nucleus, $\sim$0.04 M$_{\odot}$ yr$^{-1}$ kpc$^{-2}$ in the Bridge, and $\sim$0.01 M$_{\odot}$ yr$^{-1}$ kpc$^{-2}$ at the GRB/SN Site. Our spectroscopy across all three sites encompassed $\sim$63 kpc$^2$, or $\sim$70\% of the total host area. Taking the total SFR measured along the slit allows us to place a lower limit on the total SFR in the host of $\gtrsim$2.7 M$_{\odot}$ yr$^{-1}$. A summary of the derived ISM properties is included in Table 1.

\section{Discussion}
The association of GRB 120422A/SN 2012bz with its host galaxy is robust (see Schulze et al.\ 2012). However, it is clear from our spectra, particularly the weak emission features and SFR at the GRB/SN Site, that the explosion environment is very weakly star-forming relative to the rest of the host (unlike, for example, GRB 100316D, which was localized near the strongest star-forming region in its host; Levesque et al.\ 2011). It is worth noting that several spectral features, most notably [OII] $\lambda$3727, [OIII] $\lambda$5007, and H$\alpha$ (see Figure 2), show signs of blue-shifted or red-shifted asymmetries in emission, with these deviations from a standard line profile becoming the most pronounced at the GRB/SN Site. Such asymmetries could be indicative of outflows and inflows of ionized gas with distributed opaque clouds (e.g. Kewley et al.\ 2001), and could suggest that the host galaxy has undergone some prior merger or interaction that is still impacting the dynamics of the host regions examined here. Similar explanations have been proposed for other LGRB hosts with disturbed morphologies (e.g. Wainwright et al.\ 2007, Starling et al.\ 2011). However, it is important to note that such asymmetries can also be attributed to multiple star-forming regions or structure within the nebula (e.g. Wiersema et al.\ 2007). A proper examination of the nature of these asymmetries, and their implications for interaction activity or star-forming region components, will require higher-resolution spectroscopy and comparisons with deeper multi-band host images.

In Figure 3 we plot the existing sample of LGRB host galaxies on a luminosity-metallicity (L-Z) diagram, comparing them to contours from the L-Z relation for star-forming SDSS galaxies from Tremonti et al.\ (2004). We also plot samples of Ic-BL host galaxies, from both Modjaz et al.\ (2008, 2011) and Sanders et al.\ (2012), to examine whether there is any clear environmental distinction between Ic-BLs without LGRBs and those accompanied by LGRBs. Finally, we include data on the relativistic Ic-BL SN 2009bb from Levesque et al.\ (2010d). Data for previous LGRB hosts comes from Levesque et al.\ (2010a,b), Levesque et al.\ (2011), and references therein. From archival SDSS global photometry of the host galaxy (DR8; $g = 21.16 \pm 0.09$ and $r = 20.60 \pm 0.09$), and the corrections of Blanton \& Roweis (2007), we determine $M_B = -19.4 \pm 0.2$ for the GRB 120422A/SN 2012bz host. 

From the comparison in Figure 3, there appears to be no clear distinction in metallicity among LGRBs based on an event's classification as a subluminous burst - both subluminous LGRBs and cosmological LGRBs have an average log(O/H) + 12 = 8.2 $\pm$ 0.1. As a whole, this comparison shows that any differences within the LGRB sample at $z < 1$ cannot be discerned based on metallicities. However, this does not rule out the possibility that other burst properties (blastwave velocity, X-ray fluence, etc.) may reveal a fundamental difference in the internal properties driving the production of LGRBs in these different sub-classes. It should also be noted that these $z < 1$ GRB host studies consider only O abundances, while many single-star progenitor models depend on line-driven winds and are therefore strongly dependent on heavier element abundances which may be enhanced in GRB hosts (see Chen et al.\ 2007). Conversely, many binary progenitor models for LGRBs do not have a strong metallicity dependence (e.g. Fryer \& Heger 2005, Podsiadlowski et al.\ 2010), suggesting that a distinction between different LGRB progenitor channels may not be discernible based purely on metallicity.

The comparison between LGRB hosts and Ic-BL hosts is more complex. The Ic-BL sample from Modjaz et al.\ (2008) has an average metallicity of log(O/H) + 12 = 8.6 $\pm$ 0.1, while the Sanders et al.\ (2012) sample has a lower average metallicity of log(O/H) + 12 = 8.2 $\pm$ 0.1 (the cause of the disagreement between these two samples is not yet understood). The LGRB hosts and Sanders et al.\ (2012) Ic-BL hosts shown here also fall below the general L-Z relation for star-forming galaxies form SDSS. The physical explanation driving this offset is unclear. Mannucci et al.\ (2011) suggest that this may be attributable to a fundamental relation between metallicity, SFR, and stellar mass in star-forming galaxies, arguing that LGRBs simply occur preferentially in environments with higher SFRs. However, Kocevski \& West (2011) find that this relation is not sufficient to explain such an offset. 

From our examination of the GRB 120422A/SN 2012bz host environment, we find that this galaxy is a fairly typical LGRB host, with a low metallicity measured at both the Nucleus and the GRB/SN Site. Including this newest host within the larger sample of LGRB host galaxies, we find that there is no difference in metallicity between the subluminous and cosmological LGRB host samples. In addition, the distance of GRB 120422A/SN 2012bz from the bright star-forming nucleus of its host ($\sim$8 kpc) marks this LGRB and host environment as unique; combined with asymmetries in the emission features of the host and the lack of a clear spiral arm component in the host region, this could be indicative of prior merger or interaction activity in the host. Future work on both this and other nearby spatially-resolved LGRB hosts will allow us to further probe the nature of these galaxies. Studies of additional properties such as ionization parameter, stellar population age, and star formation history, as well as dynamical studies that can explore potential merger activity, will all be valuable in characterizing the key environmental parameters that drive progenitor formation and energetic properties for LGRBs. \\

EML is supported by NASA through Einstein Postdoctoral Fellowship grant number PF0-110075 awarded by the Chandra X-ray Center, which is operated by the Smithsonian Astrophysical Observatory for NASA under contract NAS8-03060. The Berger GRB group at Harvard is supported by the National Science Foundation under Grant AST-1107973.  Partial support was also provided by a Swift AO7 grant number 7100117.  The paper includes data gathered with the 6.5 meter Magellan Telescopes located at Las Campanas Observatory, Chile. We thank the support staff at Las Campanas for their hospitality and assistance. This paper utilized data from the Gamma-Ray Burst Coordinates Network (GCN) circulars and SDSS Data Release 8. Funding for SDSS-III has been provided by the Alfred P. Sloan Foundation, the Participating Institutions, the National Science Foundation, and the U.S. Department of Energy Office of Science. The SDSS-III web site is http://www.sdss3.org/. This work was made possible in part by collaborations and discussions at the Aspen Center for Physics, supported by NSF grant 1066293.

\begin{figure}
\center
\epsscale{1}
\plotone{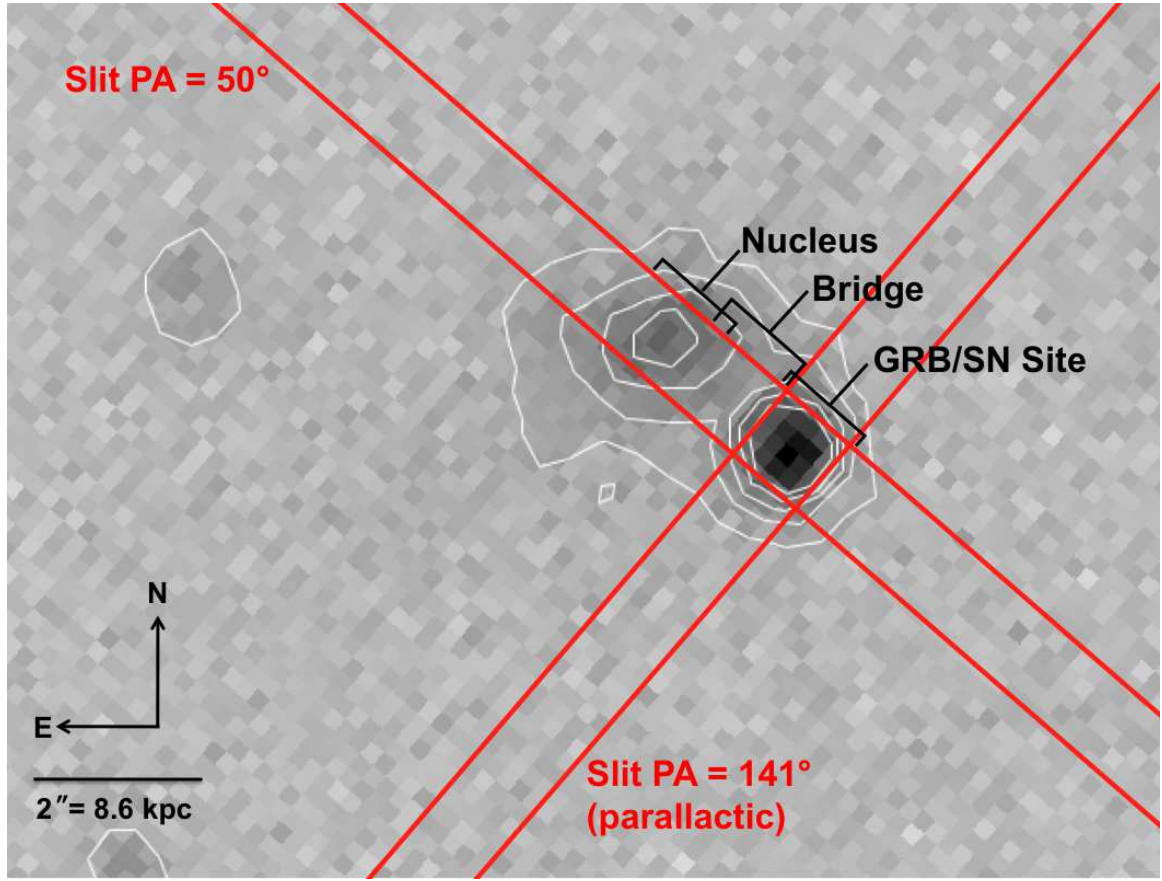}
\caption{Acquisition image showing positions of the 1$\arcsec$ slit for our observations of the GRB 120422A/SN 2012bz host galaxy. The image was taken in the r-band, with contours indicating r-band brightness; the brightness at the GRB/SN site is due to light from SN 2012bz. Regions of the host with extracted spectra are marked.}
\end{figure}

\begin{deluxetable}{l c c c c c c c c c c}
\tabletypesize{\scriptsize}
\tablewidth{0pc}
\tablenum{1}
\tablecolumns{11}
\tablecaption{\label{tab:gals} GRB 120422A Host ISM Properties}
\tablehead{
\colhead{}
&\colhead{}
&\multicolumn{5}{c}{Measured Emission Line Fluxes\tablenotemark{a}}
&\colhead{}
&\multicolumn{3}{c}{Derived ISM Properties} \\ \cline{2-7} \cline {9-11}
\colhead{Region}
&\colhead{[OII]}
&\colhead{H$\beta$}
&\colhead{[OIII]}
&\colhead{[OIII]}
&\colhead{H$\alpha$}
&\colhead{[NII]}
&\colhead{}
&\colhead{E($B-V$)\tablenotemark{b}}
&\colhead{log(O/H) + 12}
&\colhead{SFR\tablenotemark{c}} \\
\colhead{}
&\colhead{3727\AA}
&\colhead{}
&\colhead{4959\AA}
&\colhead{5007\AA}
&\colhead{}
&\colhead{6584\AA}
&\colhead{}
&\colhead{(mag)}
&\colhead{$O3N2$\tablenotemark{d}}
&\colhead{(M$_{\odot}$ yr$^{-1}$ kpc$^{-2}$)}
}
\startdata
Nucleus &$\gtrsim$8.23 &2.79$\pm$0.14 &1.92$\pm$0.10 &7.13$\pm$0.39 &7.86$\pm$0.40 &0.86$\pm$0.06 & &0.24$\pm$0.03 &8.3 $\pm$ 0.1 &$\sim$0.08 \\
Bridge &$\gtrsim$4.55 &1.40$\pm$0.10 &0.89$\pm$0.06 &2.37$\pm$0.13 &3.93$\pm$0.21 &$<$0.32 & &0.29$\pm$0.10 &$<$8.3 &$\sim$0.04 \\
GRB/SN Site &3.86$\pm$0.28 &0.45$\pm$0.04 &$<$0.29 &1.15$\pm$0.07 &1.27$\pm$0.09 &$<$0.16 & &0.31$\pm$0.13 &8.2 $\pm$ 0.1 &$\sim$0.01 \\
\enddata	      
\tablenotetext{a}{Fluxes, corrected for extinction, in units of 10$^{-16}$ ergs cm$^2$ s$^{-1}$ \AA$^{-1}$. Our analyses are dominated by the systematic errors of the environmental diagnostics.}
\tablenotetext{b}{Host E($B-V$), corrected for Galactic extinction (E($B-V$) = 0.03 mag; Schlegel et al.\ 1998).}
\tablenotetext{c}{Determined from the H$\alpha$ line fluxes and the relation of Kennicutt (1998) over an extraction region of $\sim$21 kpc$^2$.}
\tablenotetext{d}{Calibrations from Pettini \& Pagel (2004).}
\end{deluxetable}

\begin{figure}
\epsscale{1.2}
\plotone{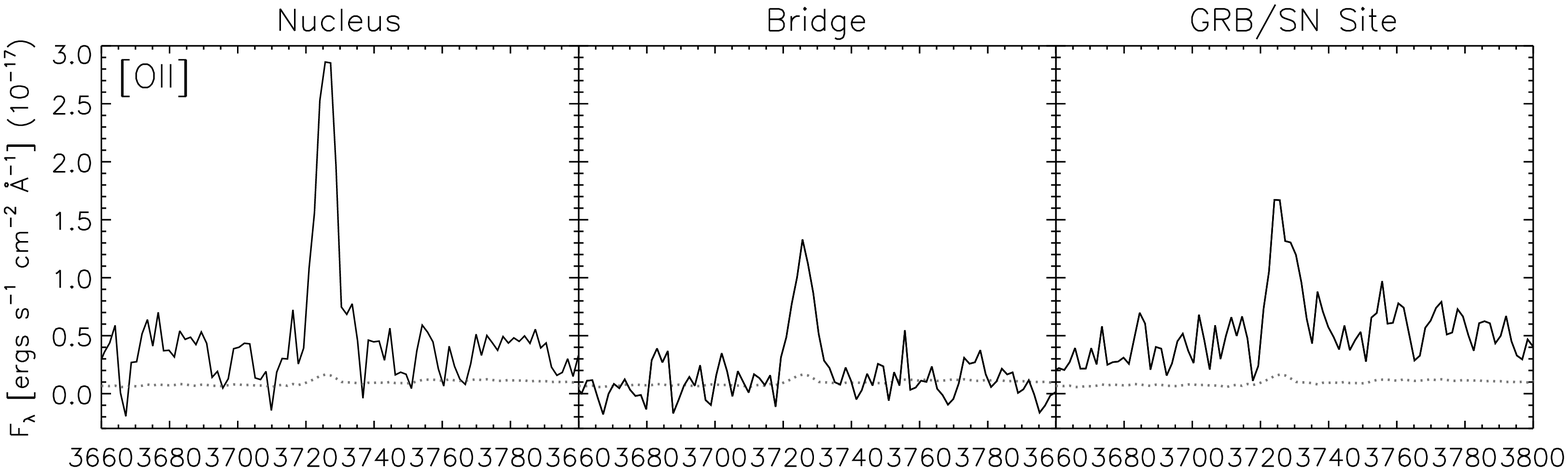}
\plotone{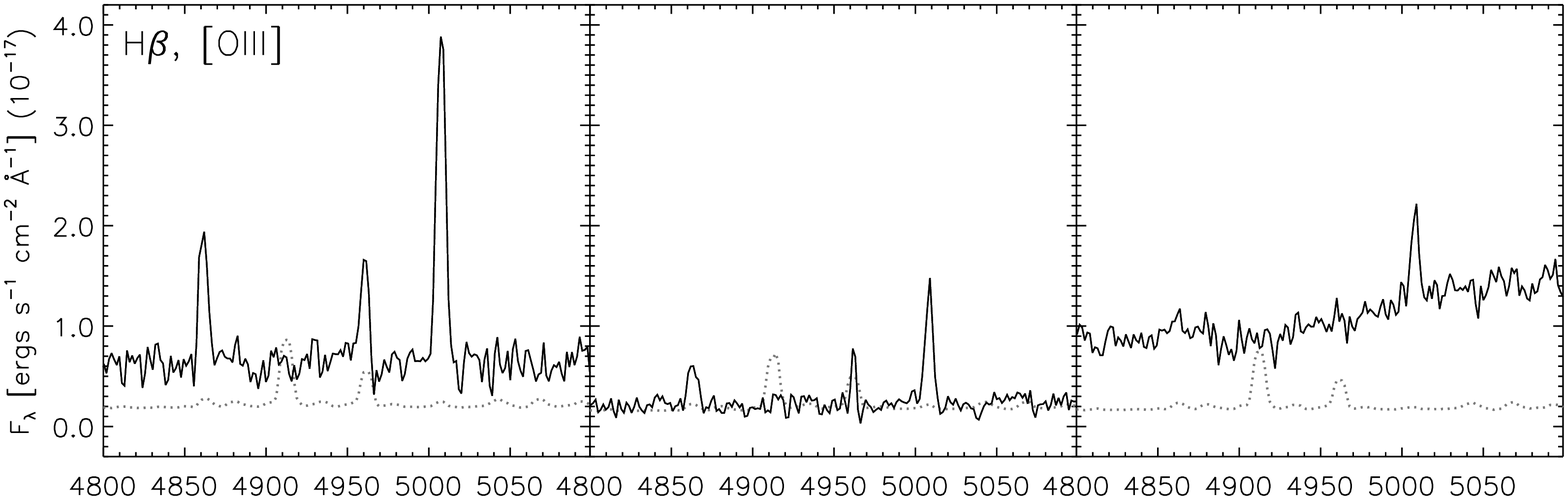}
\plotone{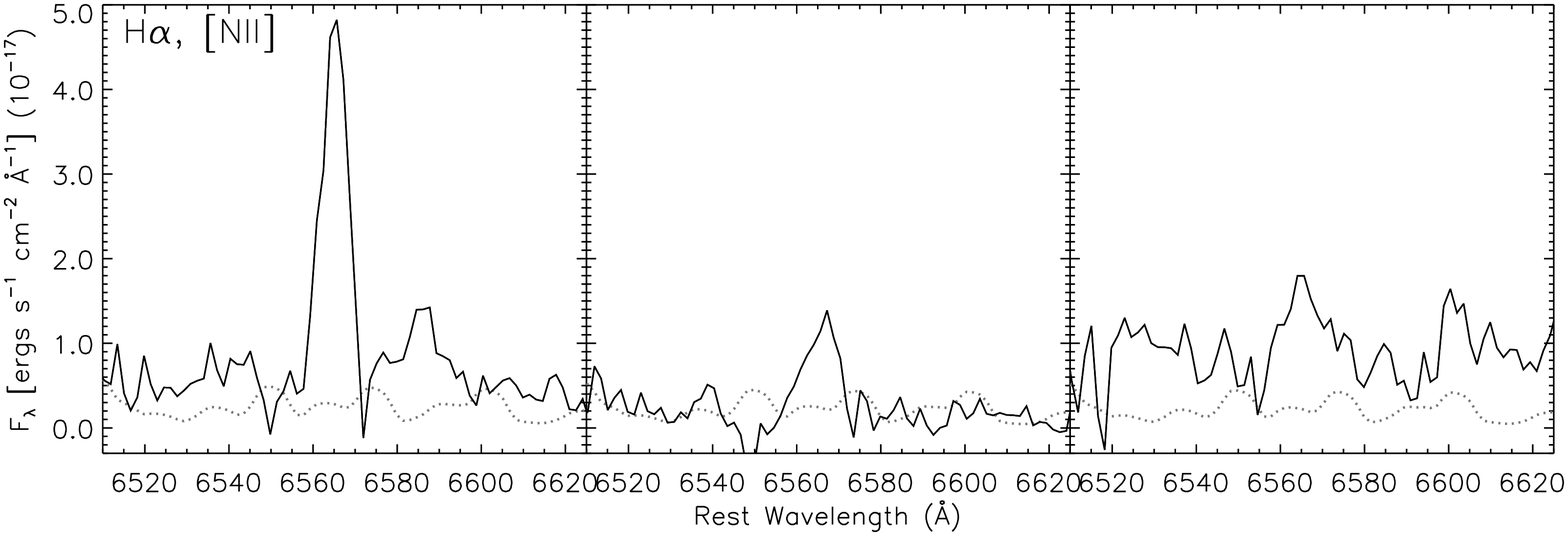}
\caption{Emission lines detected in the nucleus (left), extended emission (center), and explosion site (right) of the GRB 120422A/2012bz host galaxy. Error spectra from IRAF are overplotted as dotted gray lines. Spectra have been corrected to rest-frame wavelengths.}
\end{figure}

\begin{figure}
\epsscale{1.2}
\plotone{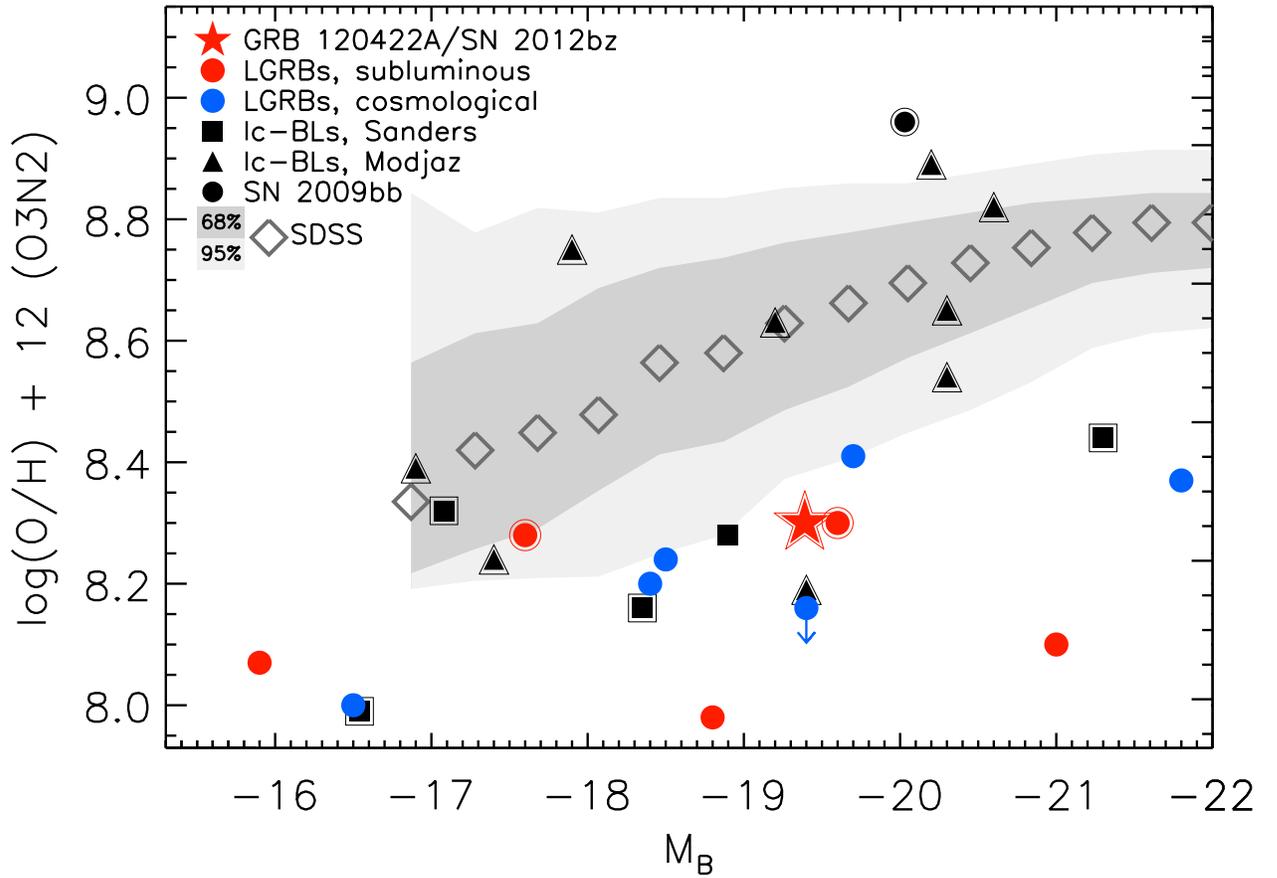}
\caption{Luminosity vs. metallicity for a sample of LGRB host galaxies, comparing ``cosmological" (blue circles), and subluminous (red circles) LGRBs; GRB 120422A is shown as a red star. All metallicities are determined by the Pettini \& Pagel (2004) {\it O3N2} metallicity diagnostic or converted to this diagnostic scale using the relations in Kewley \& Ellison (2008), and are determined based on spectroscopy from the brightest region of the host galaxy in the case of extended hosts (outlined symbols). Data for the hosts is drawn from Levesque et al.\ (2010a,b), Cano et al.\ (2011), Levesque et al.\ (2011), Vergani et al.\ (2011), and this work. We also plot the binned luminosity-metallicity data from Tremonti et al.\ (2004) for a sample of $\sim$53,000 star-forming SDSS galaxies; open diamonds represent the median in each bin, and dark and light grey regions show the contours which include 68\% and 95\% of the data respectively. Finally, we plot samples of Ic-BL host galaxies from Modjaz et al.\ (2008, 2011; black triangles) and Sanders et al.\ (2012, with absolute magnitudes from Modjaz et al.\ 2008, Modjaz et al.\ 2011, Sanders et al.\ 2011, and SDSS photometry; black squares), as well as the host of the relativistic Ic-BL SN 2009bb (black circle; Levesque et al.\ 2010d).}
\end{figure}

\end{document}